\documentclass[12pt]{article}
\pdfoutput=1
\usepackage{amssymb}
\usepackage{graphicx}
\usepackage{amsmath}
\usepackage{color}

\def\6#1{{\underline{#1}}}
\def\m6#1{{\underline{#1}\,}}

\newdimen\Tdim
\def\ispan{{\setbox0=\hbox{i}%
\Tdim\ht0\advance\Tdim\dp0\rule[-\dp0]{0pt}{\Tdim}}}
\def\jspan{{\setbox0=\hbox{j}%
\Tdim\ht0\advance\Tdim\dp0\rule[-\dp0]{0pt}{\Tdim}}}
\def\Tspan#1{{\setbox0=\hbox{#1}%
\Tdim\ht0\advance\Tdim\dp0\advance\Tdim.55ex\rule[-\dp0]{0pt}{\Tdim}\box0}}

\def\be{\begin{eqnarray}}
\def\ben{\begin{eqnarray*}}
\def\ee{\end{eqnarray}}
\def\een{\end{eqnarray*}}

\def\p{\partial}

\def\D{\mathcal{D}}

\def\=:{=\hspace{-.7em}\raisebox{1.1ex}{.}\hspace{.1em}\raisebox{-0.2ex}{.} }

\newcommand{\NF}{N_{\rm F}}
\newcommand{\NC}{N_{\rm C}}

\newcommand {\beq}{\begin{eqnarray}}
\newcommand {\eeq}{\end{eqnarray}}
\newcommand {\non}{\nonumber\\}

\def\p{\partial}
\newcommand{\vs}[1]{\vspace{#1 mm}}
\newcommand{\bpm}{\begin{pmatrix}}
\newcommand{\epm}{\end{pmatrix}}

\newcommand{\tr}{{\rm Tr}}

\newcommand{\ba}{\left( \begin{array}}
\newcommand{\ea}{\end{array} \right)}
\newcommand{\bea}{\begin{eqnarray}}
\newcommand{\eea}{\end{eqnarray}}
\newcommand{\beann}{\begin{eqnarray*}}
\newcommand{\eeann}{\end{eqnarray*}}

\makeatletter
\@addtoreset{equation}{section}

\renewcommand{\thefootnote}{\fnsymbol{footnote}}
\makeatother

\begin{document}
\begin{titlepage}
\begin{flushright}
YGHP-15-06
\end{flushright}
\begin{center}

\vs{10}
{\LARGE Semilocal Fractional Instantons}

\vs{10}
\vspace{0.5cm}
Minoru Eto${}^{1}$\footnote{meto@sci.kj.yamagata-u.ac.jp} and
Muneto Nitta${}^{2}$\footnote{nitta@phys-h.keio.ac.jp}

\vspace{0.5cm}
{\it\small 
${}^1$ Department of Physics, Yamagata University, Yamagata, 990-8560, Japan}\\
{\it\small
${}^2$ Department of Physics, and Research and Education 
Center for Natural Sciences,}\\  
{\it\small 
 Keio University, Hiyoshi 4-1-1, Yokohama, Kanagawa 223-8521, Japan}

\vspace{1cm}
{\bf Abstract}\\
\end{center}
We find semi-local fractional instantons of codimension four 
in Abelian and non-Abelian gauge theories 
coupled with scalar fields 
and the corresponding ${\mathbb C}P^{N-1}$ 
and Grassmann sigma models 
at strong gauge coupling.
They are 1/4 BPS states 
in supersymmetric theories with eight supercharges, 
carry fractional (half) instanton charges 
characterized by the fourth homotopy group $\pi_4 (G/H)$,
and have divergent energy in infinite spaces.
We construct 
exact solutions for the sigma models 
and numerical solutions for the gauge theories.
Small instanton singularity in sigma models 
is resolved at finite gauge coupling 
(for the Abelian gauge theory).  
Instantons in Abelian and 
non-Abelian gauge theories 
have negative and positive instantons charges, 
respectively,  
which are related by the Seiberg-like duality 
that changes the sign of the instanton charge.

\vspace{0.5cm}
\parbox{15cm}{
\small\hspace{15pt}
}

\end{titlepage}

\setcounter{footnote}{0}
\renewcommand{\thefootnote}{\arabic{footnote}}

\section{Introduction}

Yang-Mills instantons are solutions to
self-dual equations of pure Yang-Mills theory 
in Euclidean four space,
playing crucial roles in 
non-perturbative dynamics of 
gauge theories in four dimensions, in particular in 
supersymmetric gauge theories 
\cite{Dorey:2002ik}. 
When some scalar fields are coupled to gauge fields, 
the gauge symmetry is spontaneously broken  in the Higgs phase, 
where instantons cannot exist stably in the bulk  
as a consequence of 
the Derrick's scaling argument \cite{Derrick:1964ww}.
Instead, they can exist stably inside 
a non-Abelian vortex; 
the low-energy effective theory of 
a non-Abelian vortex in a $U(N)$ gauge theory coupled 
with $N$ fundamental scalar fields 
is the ${\mathbb C}P^{N-1}$ model in two dimensions
\cite{Hanany:2003hp,Auzzi:2003fs,Eto:2005yh},  
in which instantons in the bulk exist as 
lumps (or sigma model instantons) 
\cite{Hanany:2004ea,Eto:2004rz,Eto:2006pg}.
Such a composite configuration is 
a 1/4 BPS state 
preserving a quarter supersymmetry 
in supersymmetric theories with eight supercharges
\cite{Eto:2004rz}.
Yang-Mills instantons trapped inside a vortex as sigma model instantons 
elegantly explain a relation between 
quantum field theories in  
two and four dimensions 
\cite{Shifman:2004dr,Hanany:2004ea}.
Instanton charges also exist 
at intersections of vortices
\cite{Fujimori:2008ee}.
More general BPS composite configurations containing 
instantons in supersymmetric theories with eight supercharges
were classified in Ref.~\cite{Eto:2005sw}

In two dimensional sigma models, 
instantons are lumps 
characterized by the second homotopy group 
\cite{Polyakov:1975yp}, 
which is,  
for the ${\mathbb C}P^{N-1}$ model, 
\beq
\pi_2({\mathbb C}P^{N-1})=
\pi_2 \left( \frac{SU(N)}{SU(N-1) \times U(1)} \right)
\simeq \pi_1 \left( SU(N-1) \times U(1) \right) = {\mathbb Z}.
\label{eq:homotopy-lump}
\eeq
From the last expression of the homotopy relation, 
it is found that lumps 
can be promoted to vortices 
in gauge theories, which is, 
for the case of the ${\mathbb C}P^{N-1}$ model, 
a $U(1)$ gauge theory coupled to $N$ complex scalar fields.
Such vortices are called 
semilocal vortices \cite{Vachaspati:1991dz,Achucarro:1999it}, 
reducing to lumps in strong 
gauge coupling limit \cite{Hindmarsh:1991jq}.  
Lumps in Grassmann sigma models are promoted to 
non-Abelian vortices in a non-Abelian gauge theory 
\cite{Hanany:2004ea,Shifman:2006kd,Eto:2007yv}.
Other than vortices, possible semilocal solitons were classified 
\cite{Gibbons:1992gt,Hindmarsh:1992yy,Hindmarsh:1992ef},
including codimension-four instantons 
in quarternionic sigma models \cite{Hindmarsh:1992ef}.

In this paper, 
we construct BPS  instantons of 
codimension four
that solve a set of 1/4 BPS equations
in a $U(1)$ or $U(N)$ gauge theory coupled with scalar fields 
in four dimensions 
that reduces to 
the ${\mathbb C}P^{N-1}$ model or 
Grassmann sigma model. 
It is a Yang-Mills instanton in the gauge theory,  
and reduces to a codimension-four sigma model instanton
in the ${\mathbb C}P^{N-1}$ model 
for which we give an exact solution.  
It is characterized by the fourth homotopy group 
\beq
\pi_4({\mathbb C}P^{N-1})=
\pi_4 \left( \frac{SU(N)}{SU(N-1) \times U(1)} \right)
\simeq \pi_3 \left( SU(N-1) \times U(1) \right) = {\mathbb Z}.
\eeq
The last expression 
$\pi_3 [ SU(N-1)] \simeq {\mathbb Z}$
 denotes the homotopy group for 
Yang-Mills instantons in gauge fields, 
in parallel with Eq.~(\ref{eq:homotopy-lump}) for 
vortices, 
and so our solutions may be called 
semi-local instantons. 
Although the energy (action) of 
our solutions is divergent,
this divergence comes from 
the vortex energy while the instanton charge itself is finite.
Vortices are sheets 
linearly extending to two directions
in Euclidean four dimensions, 
having divergent energy $R^2$ with 
the system size $R$. 
Our solution is spherical and is accompanied by 
a cloud of a vortex,  giving a divergent energy 
of the same order.
Although the Derick's scaling argument 
implies the instability against shrinkage 
for scalar-field objects  
and gauge-field objects 
of codimension four, 
it is not applied to our solutions because of 
the divergent energy.  
We construct 
exact solutions for the sigma models 
and numerical solutions for the gauge theories.
The Grassmann manifold 
\beq
Gr_{N,M} = 
\frac{SU(N)}{SU(N-M) \times SU(M) \times 
U(1)}  \label{eq:Gr}
\eeq
can be reduced from either $U(M)$ gauge theory or 
$U(N-M)$ gauge theory with $N$ flavors.
The Seiberg-like duality exchanges 
the gauge groups $U(M)$ and 
$U(N-M)$ with keeping the number of flavors $N$. 
To find how the Seiberg-like duality acts 
on our solutions, we focus on the simplest case 
of the ${\mathbb C}P^2$ model;
$U(1)$ and $U(2)$ gauge theories 
reduce to the same ${\mathbb C}P^2$ model 
at strong gauge couplings. 
Our solutions are fractional instantons having 
$\pm 1/2$ fractional instanton charges 
with a minus (plus) sign for the (non-)Abelian 
gauge theory.  
We find that the Seiberg-like duality  
flips the sign of the instanton charge. 
We also find that 
a small instanton singularity in the sigma models 
is resolved at finite gauge coupling 
at least for $U(1)$ gauge theory.  
Our solution is an instanton 
for (hyper-)K\"ahler sigma models, 
while that in Ref.~\cite{Hindmarsh:1992ef} is 
for quarternionic K\"ahler sigma models. 
Another crucial difference between them 
is that our solution is BPS.  

This paper is organized as follows.
In Sec.~\ref{sec:1/4BPSeq} 
we give Lagrangian of supersymmetric $U(N)$ gauge theory,
1/4 BPS equations, 
and their solution in terms of the moduli matrix. 
In Sec.~\ref{sec:semilocal-abelian}, we present 
semi-local instanton solutions 
in the Abelian gauge theory and 
the ${\mathbb C}P^{N-1}$ sigma model.
In Sec.~\ref{sec:semilocal-na}, we present 
semi-local instanton solutions 
in the non-Abelian gauge theory and 
the Grassmannian sigma model.
Sec.~\ref{sec:summary} is devoted to summary and discussion.
In Appendix \ref{sec:duality} we describe 
the transformation of the 
topological charge under the Seiberg-like duality.

\section{The Model, BPS Equations, and Solutions}\label{sec:1/4BPSeq}
\subsection{$U(\NC)$ Gauge Theory and 1/4 BPS Equations}
In this section, we introduce 
 $U(\NC)$ gauge theory 
in (4+1)-dimensional spacetime 
with $\NF$ Higgs fields in  
the fundamental representation. 
By 
introducing additional 
$N_{\rm F}$ 
Higgs fields in the fundamental representation,
this theory can also be regarded as the bosonic part of a  
five-dimensional ${\cal N}=1$ supersymmetric $U(\NC)$ 
gauge theory with $\NF$ hypermultiplets in the fundamental 
representation. 
The fermionic part 
(and another set of $N_{\rm F}$ Higgs scalars) 
is irrelevant and is omitted 
in the following discussion.
Then the Lagrangian of the theory takes the form
\beq
{\cal L} = \tr \left[ - \frac{1}{2g^2} F_{\mu \nu} F^{\mu \nu} + \D_\mu H (\D^\mu H)^\dagger - \frac{g^2}{4} (HH^\dagger - c \mathbf 1_{\NC})^2 \right],
\label{eq:Lagrangian}
\eeq
where the Higgs fields are expressed as 
an $\NC \times \NF$ matrix $H^{rA}~(r=1,\cdots,\NC,~A=1,\cdots,\NF)$. 
The constants $g$ and $c$ are the gauge coupling constant 
and the Fayet-Iliopoulos (FI) parameter, respectively.
At the vacua (the minima of the potential) of this theory, 
the Higgs fields $H$ get  
the vacuum expectation value (vev) 
and the $U(N)$ gauge symmetry is completely broken. 
Namely the theory has only the Higgs branch due to 
the nonzero FI term. 
The moduli space of the vacua is given by a complex Grassmannian
\beq
Gr_{\NF,\NC} = \frac{SU(\NF)}{SU(\NC) \times SU(\NF-\NC) \times U(1)}.
\label{eq:vacua}
\eeq
The Abelian case $\NC=1$ corresponds to 
the projective space 
${\mathbb C}P^{\NF-1} 
\simeq {SU(\NF) / [SU(\NF-1) \times U(1)}$.
In the strong gauge coupling limit $g\to \infty$, the model reduces to 
the Grassmann sigma model 
with the target space in Eq.~(\ref{eq:vacua}).

Let us introduce the 
 complex coordinates $z$ and $w$, 
complexified gauge fields 
and the covariant derivatives by 
\be
z \equiv x^1 + ix^3,\quad
w \equiv x^2 + ix^4,\quad
\partial_z \equiv \frac{\partial_1 - i\partial_3}{2},\quad
\partial_w \equiv \frac{\partial_2 - i\partial_4}{2},\qquad\\
W_z \equiv \frac{W_1 - iW_3}{2},\quad
W_w \equiv \frac{W_2 - iW_4}{2},\quad
\D_z \equiv \frac{\D_1 - i\D_3}{2},\quad
\D_w \equiv \frac{\D_2 - i\D_4}{2},
\ee
respectively.
The 1/4 BPS equations that we consider in this paper 
are of the form 
\cite{Hanany:2004ea,Eto:2004rz}
\be
F_{13} + F_{24} = - \frac{g^2}{2}
\left( c {\bf 1}_{N_{\rm C}} - HH^\dagger \right),\quad
F_{12} = F_{34},\quad
F_{14} = F_{23},\label{vvi1}\\
\bar \D_z H = 0,\quad
\bar \D_w H = 0.\label{vvi2}
\ee
These equations can be also derived by using the 
Bogomol'nyi completion of the energy density:
\be
E &=& \int d^4x\ {\rm Tr}
\left[
\frac{1}{2g^2}F^{mn}F_{mn} 
+ \D_mH(\D^mH)^\dagger 
+ \frac{g^2}{4}
\left( c {\bf 1}_{N_{\rm C}} - HH^\dagger \right)^2 
\right]\nonumber\\
&=& \int d^4x\ {\rm Tr}
\bigg[
\frac{1}{g^2}\bigg\{\left(F_{13} + F_{24} + \frac{g^2}{2}
\left( c {\bf 1}_{N_{\rm C}} - HH^\dagger \right)\right)^2
+ \left(F_{12} - F_{34}\right)^2 
\nonumber\\
&&
+ \left(F_{14} - F_{23}\right)^2 
\bigg\} + 4 \bar \D_z H (\bar \D_z H)^\dagger
+ 4 \bar \D_w H (\bar \D_w H)^\dagger
\nonumber\\
&&
+ i\left\{\D_1H(\D_3H)^\dagger - \D_3H(\D_1H)^\dagger\right\}
+ i\left\{\D_2H(\D_4H)^\dagger - \D_4H(\D_2H)^\dagger\right\}
\nonumber\\
&& - \left(F_{13} + F_{24}\right)
\left( c {\bf 1}_{N_{\rm C}} - HH^\dagger \right) + \frac{1}{2g^2}F_{mn}\tilde F^{mn}
\bigg]\nonumber\\
&=& \int d^4x\ {\rm Tr}
\bigg[
\frac{1}{g^2}\bigg\{\left(F_{13} + F_{24} + \frac{g^2}{2}
\left( c {\bf 1}_{N_{\rm C}} - HH^\dagger \right)\right)^2
+ \left(F_{12} - F_{34}\right)^2 
\nonumber\\
&&
+ \left(F_{14} - F_{23}\right)^2 
\bigg\} + 4 \bar \D_z H (\bar \D_z H)^\dagger
+ 4 \bar \D_w H (\bar \D_w H)^\dagger
\nonumber\\
&&
- c \left(F_{13} + F_{24}\right) + \frac{1}{g^2}F_{mn}\tilde F^{mn}
+ \partial^mJ_m\bigg],
\label{eq:bogo}
\ee
where we have defined the current
\be
J_1 \equiv iH\D_3H^\dagger ,\quad
J_3 \equiv -iH\D_1H^\dagger ,\non
J_2 \equiv iH\D_4H^\dagger ,\quad
J_4 \equiv -iH\D_2H^\dagger ,
\ee
and have used the following identity
\be
{\rm Tr}\left[
\partial_1 J_1 + \partial_3 J_3\right]
&=& i {\rm Tr}\big[
\D_1H\D_3H^\dagger - \D_3H\D_1H^\dagger - iHH^\dagger F_{13}
\big],\non
{\rm Tr}\left[
\partial_2 J_2 + \partial_4 J_4\right]
&=& i {\rm Tr}\big[
\D_2H\D_4H^\dagger - \D_4H\D_2H^\dagger - iHH^\dagger F_{24}
\big].
\ee

We thus have found the energy bound from below
\be
E \ge \int dx^4\ {\rm Tr}
\left( V_z + V_w + {\cal I} + \partial^mJ_m\right),
\ee
where the bound consists of three parts
\be
V_z = - c{\rm Tr}(F_{13}),\quad
V_w = - c{\rm Tr}(F_{24}),\quad
{\cal I} = \frac{1}{g^2}{\rm Tr}\left(F_{mn}\tilde F^{mn}\right).
\ee
It is saturated when the 1/4 BPS equations (\ref{vvi1}) and (\ref{vvi2})
are satisfied. 
Using the BPS equations, the current is expressed by
\be
J_m = \frac{1}{2}\partial_m {\rm Tr}\left[ HH^\dagger\right].
\ee
Under the spacial integrations, we have
\be
\int d^4x\ {\rm Tr} V_z &=& \int dx^2dx^4 \int dx^1dx^3\ (-c {\rm Tr}F_{13}) = -2\pi c k_z S_w,\\
\int d^4x\ {\rm Tr} V_w &=& \int dx^1dx^3 \int dx^2dx^4\ (-c {\rm Tr}F_{24}) = -2\pi c k_w S_z,
\ee
where we introduce {\it negative} integers
\be
k_{z} = \frac{1}{2\pi} \int dx^1dx^3\ {\rm Tr}F_{13},\qquad
k_{w} = \frac{1}{2\pi} \int dx^2dx^4\ {\rm Tr}F_{24},
\ee
and the infinite areas $S_{z} = \int dx^1dx^3$ and $S_w = \int dx^2dx^4$.
We have another topological number, the instanton charge
\be
I = \frac{g^2}{4\pi^2}\int dx^4\ {\cal I}.
\ee
Furthermore, $J_m$ goes exponentially rapidly to zero at spacial infinity, leading to $\int dx^4\ \p_m J_m = 0$. In summary, the 1/4 BPS solution has mass
\be
E = -2\pi c (k_z S_w+ k_w S_z) + \frac{4\pi^2}{g^2} I.
\ee
Note that $k_{z,w}$ is always negative due to our choice of sign when we performed 
the Bogomol'nyi completion at Eq.~(\ref{eq:bogo}). Therefore, the first two terms,
the tension of the vortices, are positive definite. On the other hand, the instanton number
$I$ is not restricted to be positive or negative. Below, we will show that $I$ is negative
in the Abelian gauge theory while it is positive in the non-Abelian gauge theory.
When $I$ is negative, it may be suitable that the instanton charge give a sort of binding 
energy, rather than a particle.

\subsection{Solving 1/4 BPS Equations 
in Terms of the Moduli Matrix}
Let us solve the BPS equations (\ref{vvi1}) and (\ref{vvi2}).
Eq.~(\ref{vvi2}) can be easily solved by introducing  the $\NC \times \NF$ moduli
matrix whose components are holomorphic in both $z$ and $w$ \cite{Eto:2004rz,Eto:2004rz}
\be
H = S^{-1} H_0(z,w),\quad
\bar W_z = -iS^{-1}\bar\partial_z S,\quad
\bar W_w = -iS^{-1}\bar\partial_w S,
\ee
where $S \in GL(\NC,{\mathbb C})$.
The last two equations can be rewritten by
\be
\bar\D_z S^{-1} = 0,\quad
\bar\D_w S^{-1} = 0.
\ee
The last two equations\footnote{
The following useful indenties hold
\be
\left[\bar\D_z,\bar\D_w\right] &=& 
\frac{i}{4}
\left[
F_{12} - F_{34} + i (F_{14} - F_{23})\right],\label{bd,bd'}\\
\left[\bar\D_z,\D_z\right] &=& 
\frac{1}{2}F_{13},\\
\left[\bar\D_w,\D_w\right] &=& 
\frac{1}{2}F_{24},\\
\left[\bar\D_z,\D_w\right] &=& 
\frac{i}{4}
\left[
F_{12} + F_{34} - i (F_{14} + F_{23})\right].\label{bd,d'}
\ee
}  in Eq.~(\ref{vvi1}) are automatically satisfied and insure the
integrability of two operators $\bar\D_z$ and $\bar\D_w$
\be
\left[\bar\D_z,\bar\D_w\right] = 
\frac{i}{4}
\left[
F_{12} - F_{34} + i (F_{14} - F_{23})\right] = 0.
\ee
The last unsolved equation is the first one of (\ref{vvi1}).
Let us rewrite it by using a gauge invariant $\NC\times\NC$
hermite matrix 
\be
\Omega \equiv SS^\dagger.
\ee
By using
$\D S^{-1} = -S^{-1}(\partial\Omega)\Omega^{-1}$, 
$\D' S^{-1} = -S^{-1}(\partial'\Omega)\Omega^{-1}$
and the equations (\ref{bd,bd'})--(\ref{bd,d'}),
the field strength $F_{mn}$ can be expressed in terms of $\Omega$ as
\be
F_{12} + F_{34} - i (F_{14} + F_{23}) &=& 4iS^{-1}\bar\partial\left[
(\partial'\Omega)\Omega^{-1}\right] S,\\
F_{12} + F_{34} + i (F_{14} + F_{23}) &=& -4iS^{-1}\bar\partial'\left[
(\partial\Omega)\Omega^{-1}\right] S,\\
F_{13} &=& - 2S^{-1}\bar\partial\left[(\partial\Omega)\Omega^{-1}\right]S,\\
F_{24} &=& - 2S^{-1}\bar\partial'\left[(\partial'\Omega)\Omega^{-1}\right]S,
\ee
where $
F_{34} = F_{12},\ 
F_{23} = F_{14}
$ from the BPS solution.
Then the last of the BPS equation can be expressed as
\be
\bar\partial_z\left[(\partial_z\Omega)\Omega^{-1}\right]
+ \bar\partial_w\left[(\partial_w\Omega)\Omega^{-1}\right]
= \frac{cg^2}{4} \left[{\bf 1}_{\NC} - \Omega_0\Omega^{-1}\right],
\label{master_vvi}
\ee
that we call the master equation,
with 
\be
\Omega_0 \equiv c^{-1} H_0H_0^\dagger.
\ee
The energy density can be expressed as follows
\be
t_{\rm v} &=& V_z + V_w 
= \frac{c}{2}\triangle_4\log\det\Omega,\\
{\cal I} &=&
\frac{1}{g^2}{\rm Tr}\left(
F_{12}F_{34} + F_{14}F_{23} - F_{13}F_{24}
\right)\nonumber\\
&=& \frac{4}{g^2}{\rm Tr}\bigg[
\bar\partial_z \left\{(\partial_w \Omega)\Omega^{-1}\right\}
\bar\partial_w \left\{(\partial_z \Omega)\Omega^{-1}\right\}
-\ \bar\partial_z \left\{(\partial_z \Omega)\Omega^{-1}\right\}
\bar\partial_w \left\{(\partial_w \Omega)\Omega^{-1}\right\}
\bigg].
\ee

\section{Semilocal Instantons in Abelian Gauge Theory and 
the ${\mathbb C}P^{N-1}$ Model}\label{sec:semilocal-abelian}

\subsection{Single spherical solution}

As the minimal model admitting a semi-local instanton, 
let us consider $U(1)$ gauge theory with $\NF=3$ 
charged Higgs fields.

The moduli matrix for a spherically symmetric solution 
in the ${\mathbb C}P^2$ model 
is given by
\beq
H_0(z,w) &=& \sqrt{c}\, (z,\ w,\ a),
\label{eq:mm_1inst}
\eeq
where $a$ represent size and phase moduli. 
This yields 
\beq
\Omega_0 &=& r^2 + |a|^2 ,
\label{eq:omega0_1inst}
\eeq
where we have defined complex coordinates by
\be
z = r e^{i\eta} \cos \xi ,\quad w = r e^{i\lambda} \sin \xi .
\label{eq:complex_coordinate}
\ee
with 
$r \in [0,\infty),\ \xi \in [0,\pi/2],\ \eta \in [0,2\pi),\ \lambda \in [0,2\pi)$.
Since the source $\Omega_0$ is a function of $r$ only, the master equation (\ref{master_vvi})
can be reduced into the following ordinary differential equation for $\Omega(r) = e^{u(r)}$,
\begin{eqnarray}
u'' + \frac{3 u' }{r} - g^2c \left[ 1 - (r^2 + |a|^2)e^{-u}\right]  = 0.
\label{eq:master_axial}
\end{eqnarray} 
It is worth to point out that 
if we replace the coefficient 3 in the second term by 1, 
we obtain the master equation for 
a BPS semi-local vortex in the Abelian-Higgs 
model with $\NF\ge 2$ in $2+1$ dimensions. 

\begin{figure}[t]
\begin{center}
\includegraphics[width=10cm]{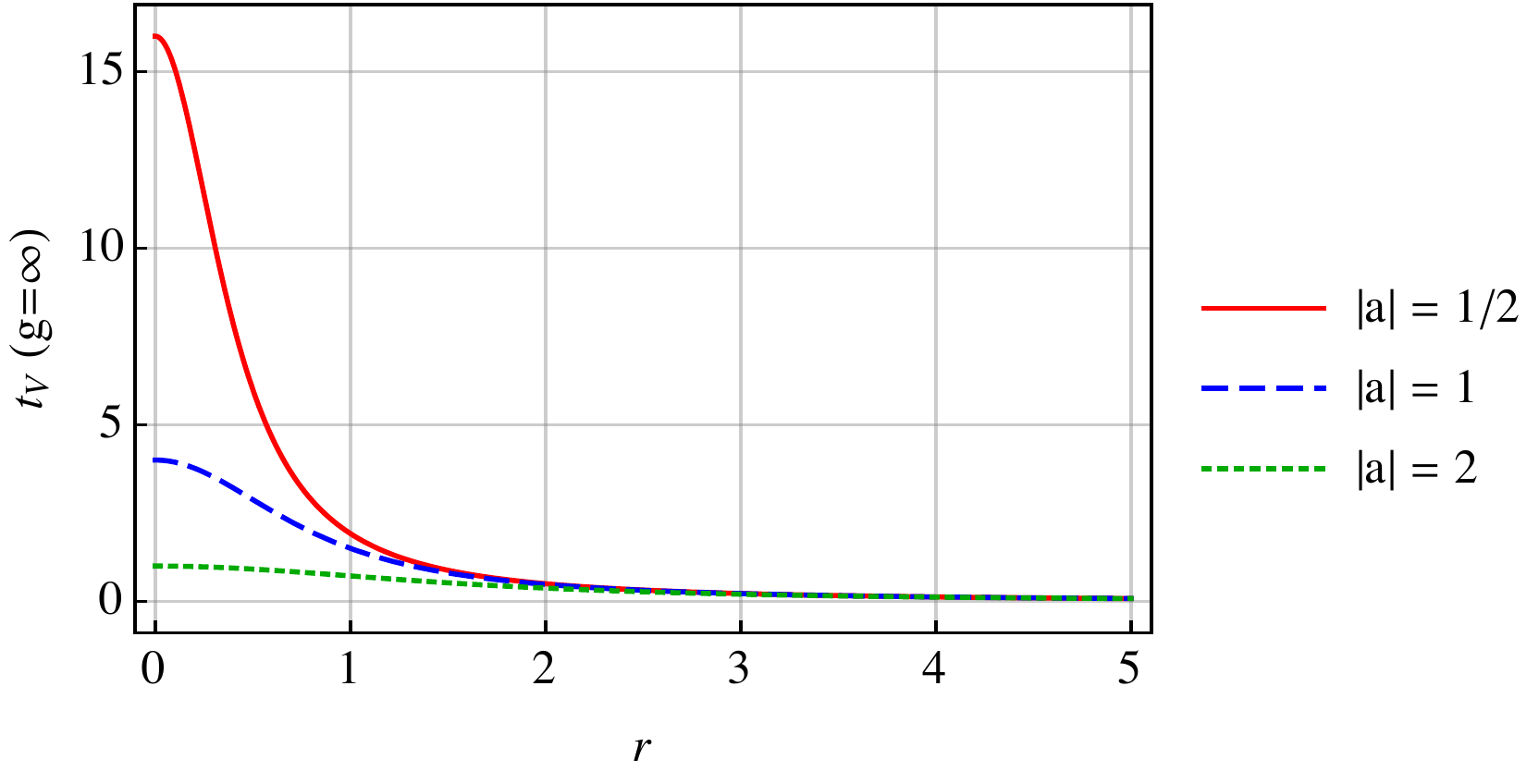}
\caption{Energy density $t_{\rm v}$ of one instanton 
with a size modulus $|a|=1/2,1,2$ 
at the strong gauge coupling (sigma model) limit $g \to \infty$.}
\label{tv_1ins}
\end{center}
\end{figure}

In the sigma model limit $g\to \infty$, 
the energy density 
coincides with the vortex charge density, 
because the contribution to the energy density from 
the instanton charge density vanishes.
At the same time we have the exact solution $\Omega = \Omega_0$ to 
the master equation (\ref{master_vvi}).
The energy density is then given by
\begin{eqnarray}
{\cal E}_{g \to \infty} = 
t_{\rm v} = \frac{c}{2}\left(\p_r^2 + \frac{3}{r}\p_r\right) \log \Omega
= \frac{2c(r^2 + 2|a|^2)}{(r^2 + |a|^2)^2}.
\end{eqnarray}
The energy density  $t_{\rm v}$ is 
shown in Fig.~\ref{tv_1ins}. 
The integrated energy is quadratically divergent 
$E = \int d^4 x\, t_{\rm v} \sim R^2$
with the size $R$ of the system. 
The quadratic divergence exists when 
a vortex is sheet-like, 
but the same divergence still exists 
for the spherical configuration 
that we have constructed.
This may be understood as a cloud of a vortex.

\begin{figure}[ht]
\begin{center}
\includegraphics[height=7cm]{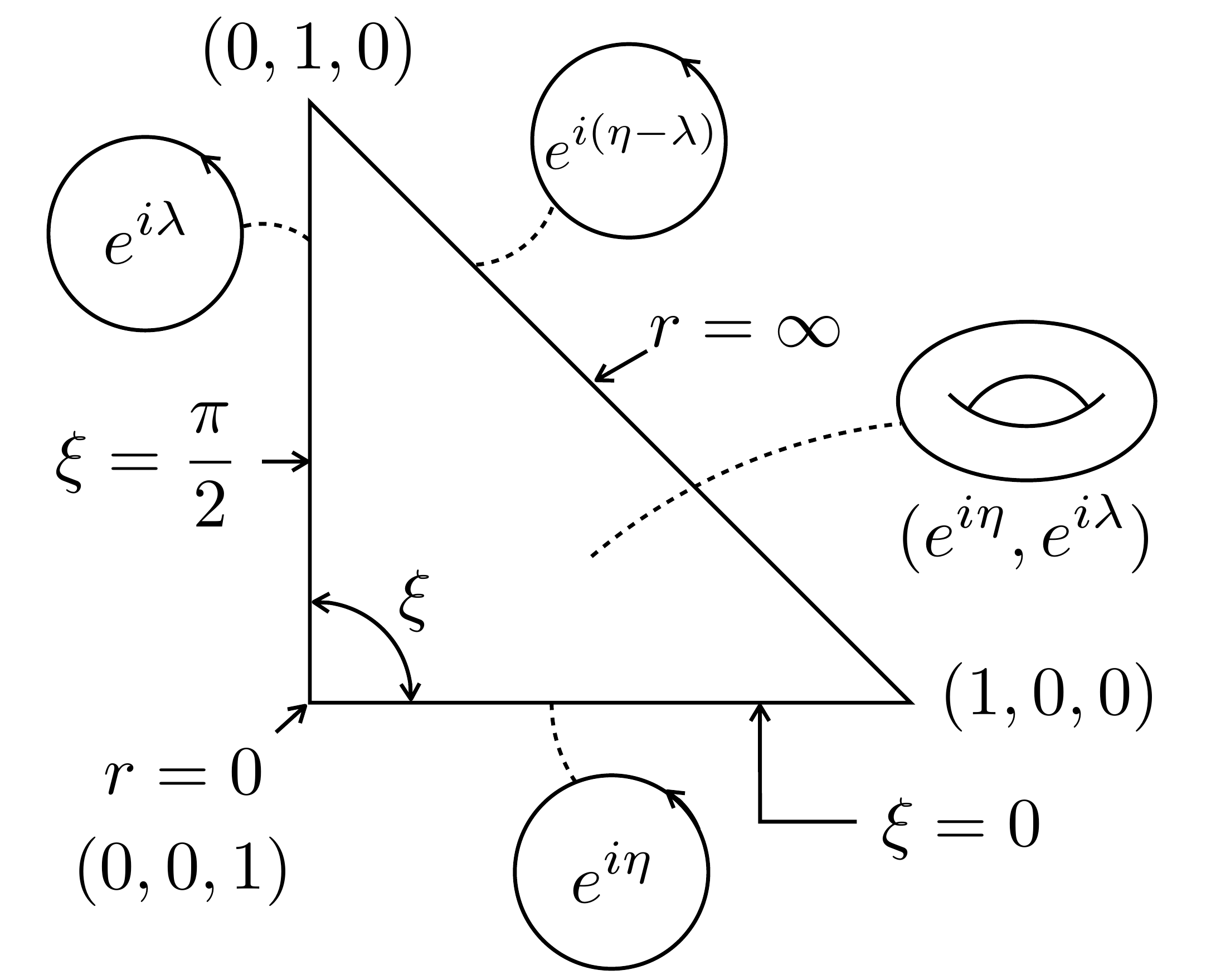}
\caption{The map from $\mathbb{R}^4$ to $\mathbb{C}P^2$.
The toric diagram is shown, for which 
the horizontal and vertical axes denote 
$|H_1|^2$ and $|H_2|^2$, respectively.
}
\label{fig:toric_cp2}
\end{center}
\end{figure}

The profile functions of the Higgs fields for the exact solution at $g \to \infty$ gives a map
from $\mathbb{R}^4$ to $\mathbb{C}P^2$
\be
H(r,\xi,\eta,\lambda) = \sqrt c \left(
\begin{array}{ccc}
\dfrac{r e^{i\eta}\cos\xi}{\sqrt{r^2 + |a|^2}}, & 
\dfrac{r e^{i\lambda}\sin\xi}{\sqrt{r^2 + |a|^2}}, & 
\dfrac{a}{\sqrt{r^2 + |a|^2}}
\end{array}
\right),
\label{eq:sol_H_Abelian}
\ee
where the overall phase is gauged away. Fig.~\ref{fig:toric_cp2} shows the map
onto the topic diagram of $\mathbb{C}P^2$.
At large distance $r\to \infty$, the configuration in Eq.~(\ref{eq:sol_H_Abelian}) reduces to
\be
H(r\to \infty,\xi,\eta,\lambda) 
= \sqrt c \left(
\begin{array}{ccc}
e^{i\eta}\cos\xi, & e^{i\lambda}\sin\xi ,& 0
\end{array}
\right) 
\sim  
 \sqrt c \left(
\begin{array}{ccc}
e^{i\eta-i\lambda}\cos\xi, & \sin\xi, & 0
\end{array}
\right) 
\label{eq:sol_H_Abelian-infty}
\ee
where $\sim$ denotes a gauge equivalence.
Therefore, 
the boundary $S^3$ at the spacial infinity of $\mathbb{R}^4$ is mapped to
a submanifold $\mathbb{C}P^1 \simeq S^2$ corresponding to the diagonal edge in Fig.~\ref{fig:toric_cp2}.
Our configuration therefore has 
additional topological charge,
characterized by the Hopf map\footnote{A soliton in four Euclidean space 
characterized by the same Hopf map at the 
boundary $S^3$ 
was studied before in an $SU(2)$ gauge theory with triplet Higgs field, by which the $SU(2)$ gauge symmetry 
is spontaneously broken to 
a $U(1)$ subgroup \cite{He:2014eoa}. 
Our soliton may be understood as a Higgsed version of it.
}
\beq 
 \pi_3(S^2) \simeq \mathbb{Z}  \label{eq:Hopf}.
\eeq

\begin{figure}[t]
\begin{center}
\includegraphics[width=11cm]{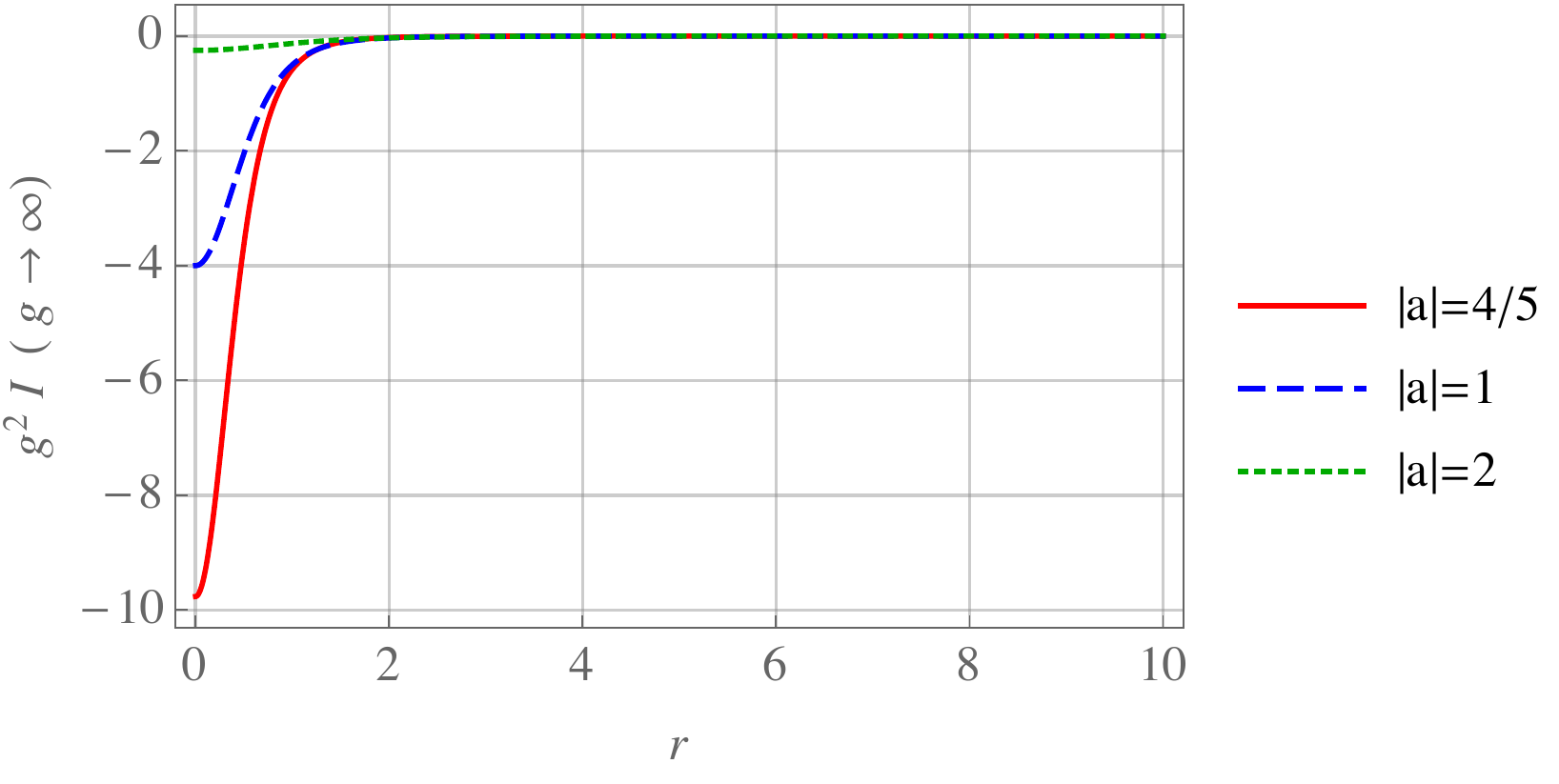}
\caption{Instanton charge density $g^2{\cal I}$ 
of a single instanton 
with a size modulus $|a|=4/5,1,2$ 
at the strong gauge coupling (sigma model) limit $g \to \infty$.}
\label{fig:abelian_inst_density_infinite}
\ \\
\includegraphics[width=10.5cm]{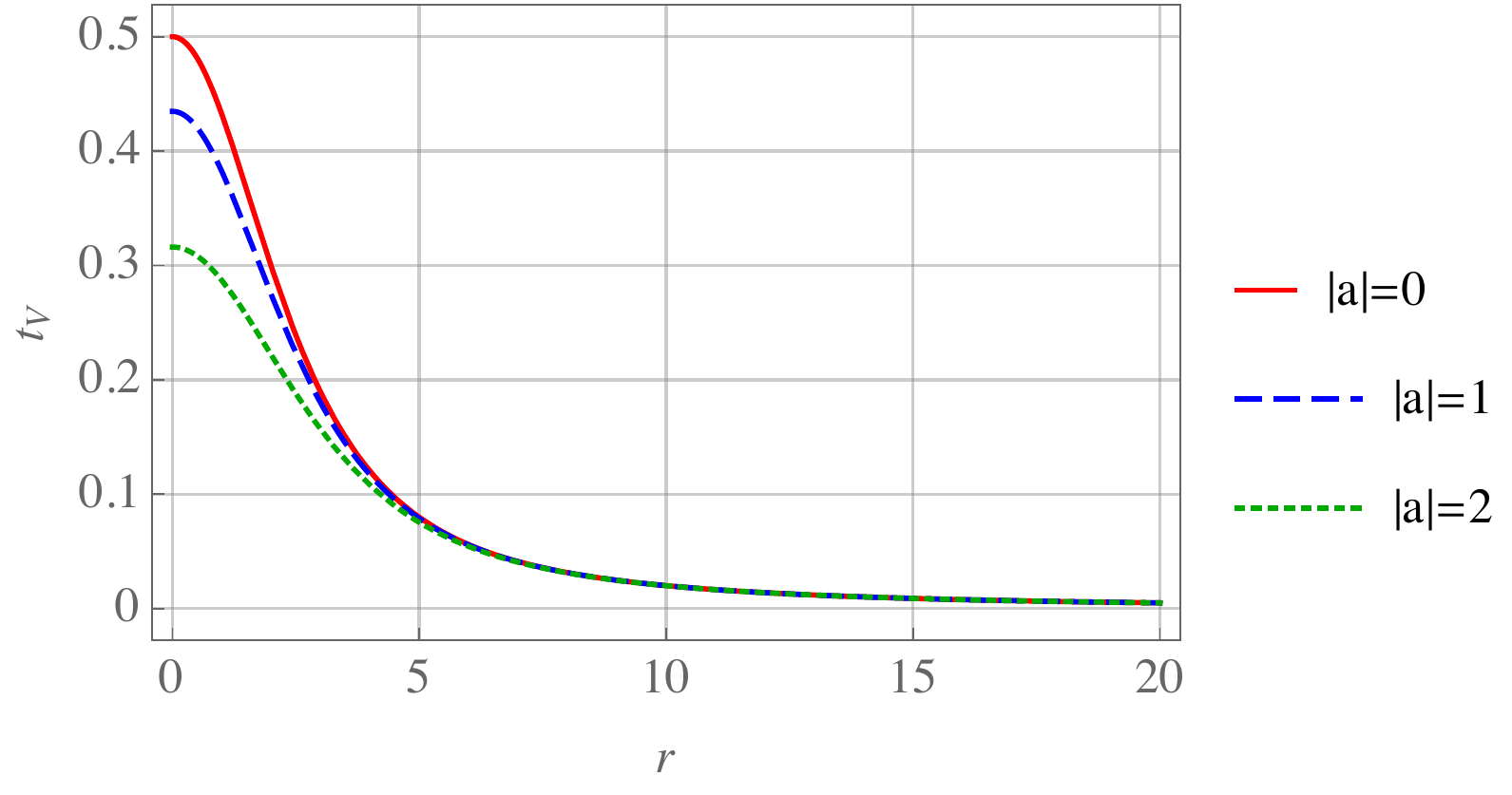}
\caption{$t_{\rm v}$ of a single instanton 
with a size modulus $|a|=0,1,2$ 
at the finite gauge coupling $g^2c  = 1$.}
\label{fig:tv_1ins_finite}
\end{center}
\end{figure}

Although the instanton charge density at $g \to \infty$ does not contribute to the 
total energy density,  
$g^2 {\cal I}$ has a nonzero support around the origin as 
is given by
\begin{eqnarray}
g^2 {\cal I} = -\frac{4 |a|^2}{ \left(r ^2+|a|^2\right)^3}.
\label{eq:one-instanton-charge}
\end{eqnarray}
The instanton charge density is shown in Fig.~\ref{fig:abelian_inst_density_infinite}.
The instanton charge of this configuration is finite and interestingly fractional as
\begin{eqnarray}
I = \frac{1}{4\pi^2} \int d^4x \ g^2{\cal I}
= - \frac{1}{2}.
\end{eqnarray}
Thus, this is a fractional instanton, 
or a {\it meron}.

The solution at $g\to\infty$ has a small instanton singularity at $a =0$ where
the energy density diverges and 
the instanton charge density becomes a delta function.
The small instanton singularity is resolved for the finite gauge coupling constant $g$,
because the typical mass scale $g\sqrt c$ turns into the theory. 
This is a good property for $g < \infty$ but we have to pay
the cost that the master equation (\ref{eq:master_axial}) cannot  be solved analytically anymore.

\begin{figure}[t]
\begin{center}
\includegraphics[width=10cm]{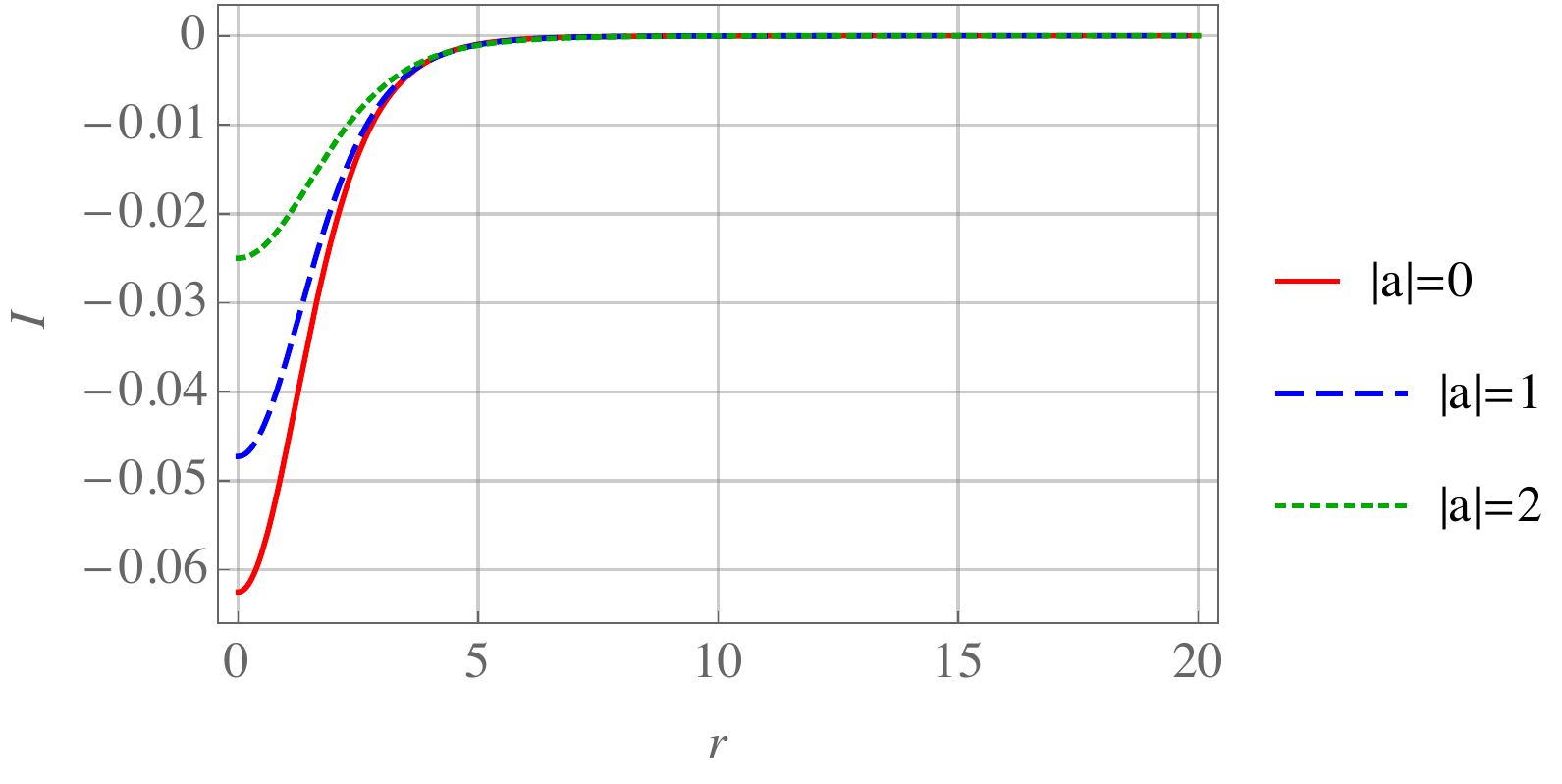}
\caption{${\cal I}$ of the 1 instanton for the finite gauge coupling $g^2c  = 1$.}
\label{fig:I_1ins_finite}
\end{center}
\end{figure}

Here we numerically solve the master equation 
for finite gauge couplings. 
In Fig.~\ref{fig:tv_1ins_finite}, we show numerical solutions
with $|a| = 0,1,2$ as examples. 
Even for $a=0$, no singular behaviors are observed 
so that the small instanton singularity is resolved.
Compared these with the configurations in Fig.~\ref{tv_1ins}, it is seen that
the energy density distributions
of the finite gauge coupling constant tend to be 
broader and more smeared.
We also show the {\it negative} instanton charge densities in Fig.~\ref{fig:I_1ins_finite}.
Compared to the infinite gauge coupling limit shown in Fig.~\ref{fig:abelian_inst_density_infinite},
the peaks of instanton charge densities become very small; They are about 1 \%. 
Remarkably, the instanton charge contribution to the total energy density does not
vanish even for $a=0$. 
Note that setting $a=0$ in Eq.~(\ref{eq:mm_1inst}) implies that the third component of the
the Higgs field is 0 everywhere.
Namely, the solution for $a=0$ remains a solution for the theory 
with $\NF=2$ for which the moduli space of vacua is $\mathbb{C}P^1 \simeq S^2$. 

\subsection{Multiple instantons}
\begin{figure}[ht]
\begin{center}
\includegraphics[width=6cm]{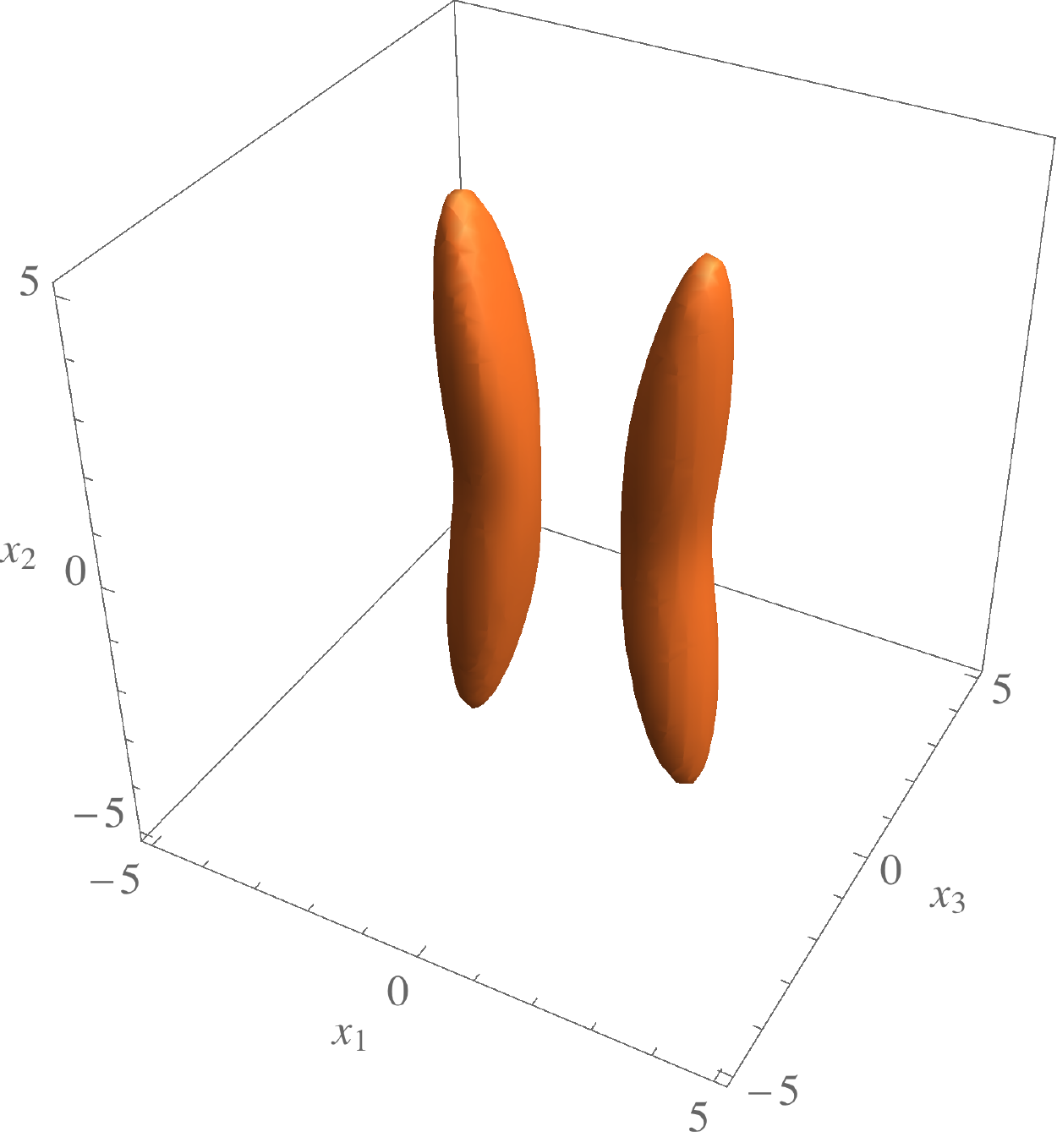}
\caption{Sufaces on which $t_{\rm v} = 1$ for 2 instantons in $x^{1,2,3}$ space
at $x^4=0$. We take $c=1$ and $g\to\infty$.}
\label{prof_2ins}
\end{center}
\end{figure}

Non-spherical multiple instantons 
can be constructed in the ${\mathbb C}P^2$ model by
the moduli matrix of the form of 
\begin{eqnarray}
H_0 = (z^m + \cdots , w^n + \cdots, a)
\end{eqnarray}
where $\cdots$ denote polynomials with degrees 
$m$ of $z$ and $n$ of $w$.
This gives 
\begin{eqnarray}
\Omega = |z^m + \cdots|^2 + |w^n +\cdots|^2 + a^2.
\end{eqnarray}
For instance, the configuration 
given by the moduli matrix
\begin{eqnarray}
H_0 = ( (z+1.5)(z-1.5),\ w,\, 1)
\end{eqnarray}
is shown in Fig.~\ref{prof_2ins}.

If we go to the ${\mathbb C}P^3$ model,
another spherically symmetric 
configuration 
made of four instantons can be constructed.
Let us consider the moduli matrix given by
\begin{eqnarray}
H_0 = ( z^2, w^2, \sqrt 2 zw, a).
\end{eqnarray}
This yields 
\begin{eqnarray}
\Omega = r^4 + |a|^2.
\end{eqnarray}
The vortex and instanton charge densities  are given by
\begin{eqnarray}
t_{\rm v} = \frac{4r^2(r^4+3|a|^2)}{(r^4+|a|^2)^2},\quad
g^2 {\cal I} = -\frac{32 |a|^2 \rho ^4}{\left(r^4+|a|^2\right)^3},
\end{eqnarray}
respectively. 
The integrated energy is again quadratically 
divergent $E \sim R^2$ with the system size $R$.
As before, although the energy is divergent, 
the total instanton charge is finite:
\begin{eqnarray}
I = \frac{g^2}{4\pi^2} \int d^4x\ {\cal I} = -2 
\end{eqnarray}
that is four multiple of 1/2 instantons.

\section{Semilocal Instantons in Non-Abelian Gauge Theory 
and the Grassmann Sigma Model}\label{sec:semilocal-na}

\subsection{Single spherical solution}

As the simplest case, we consider $U(2)$ gauge theory 
with $\NF=3$ flavors.
The moduli space of vacua is
\beq
{\cal M}^{U(2)}_{\NF=3} = \frac{SU(3)_{\rm F}}{SU(2)_{\rm C+F} \times U(1)_{\rm C+F}}
\simeq {\mathbb C}P^2
\eeq
which is the same with that of the $U(1)$ gauge theory with $\NF=3$ flavors 
because of the Seiberg-like duality $\NC \leftrightarrow \NF - \NC$.
In the strong gauge coupling limit, these two models 
reduce to the same ${\mathbb C}P^2$ model.

The moduli matrix of a 
spherically symmetric solution is given by
\beq
H_0(z,w) = \sqrt c\,
\left(
\begin{array}{ccc}
1 & \dfrac{w}{a} & 0 \\
0 & z & a
\end{array}
\right),
\label{eq:mm_na_radial}
\eeq
that gives
\beq
\Omega_0 =  \left(
\begin{array}{cc}
1 + \left|\dfrac{w}{a}\right|^2 & \dfrac{z^* w}{a}\\
\dfrac{z w^*}{a^*} & |a|^2 + |z|^2
\end{array}
\right),\quad
\det \Omega_0 = r^2 + |a|^2.
\eeq
This is similar to $\Omega_0$ which we encountered in the $U(1)$ gauge theory.

In the strong gauge coupling limit, the master equation is again exactly solved by
$\Omega = \Omega_0^{-1}$. The profile functions of the Higgs fields are given by
\be
S^{-1} &=& \frac{1}{\sqrt{1+|z|^2+|w|^2}} \left(
\begin{array}{cc}
\sqrt{1+|z|^2} & - \frac{z^*w}{\sqrt{1+|z|^2}}\\
0 & \sqrt{\frac{1+|z|^2+|w|^2}{1+|z|^2}}
\end{array}
\right),\\
H &=& \left(
\begin{array}{ccc}
\sqrt{\frac{1+|z|^2}{1+|z|^2+|w|^2}} & \frac{w}{\sqrt{(1+|z|^2)(1+|z|^2+|w|^2)}} & - \frac{z^*w}{\sqrt{(1+|z|^2)(1+|z|^2+|w|^2)}}\\
0 & \frac{z}{\sqrt{1+|z|^2}} & \frac{1}{\sqrt{1+|z|^2}}
\end{array}
\right).
\ee
Consider
$SU(2)_C$ invariant quantities, the determinants of 2 by 2 submatrices taking $i$ and $j$-th column from
$H$, are given by
\be
\det H_{12} = \frac{z}{\sqrt{1+|z|^2+|w|^2}},\ 
\det H_{23} = \frac{w}{\sqrt{1+|z|^2+|w|^2}},\ 
\det H_{13} = \frac{1}{\sqrt{1+|z|^2+|w|^2}}.
\ee
This is exactly identical to the solution in the Abelian theory given in Eq.~(\ref{eq:sol_H_Abelian}).
From these we can calculate the instanton density
\beq
g^2 {\cal I} = 
\frac{4a^2}{(r^2 + a^2)^3},
\eeq
that coincides with the one of 
Eq.~(\ref{eq:one-instanton-charge}) 
with opposite sign. Thus, we have a positive contribution $I$ to the energy density,
\be
I = \frac{1}{4\pi^2} \int dx^4\ g^2{\cal I} = \frac{1}{2}.
\ee
The sign flip between the instanton charges in
the original theory with $\NC$ 
and the dual theory 
$\NC = \NF - \NC$
is shown in Appendix \ref{sec:duality}.

On the other hand, we numerically solve the master equation for the finite gauge coupling constant.
For simplicity, we will consider the moduli matrix (\ref{eq:mm_na_radial}) with $a=1$.
In terms of the complex coordinate coordinates in Eq.~(\ref{eq:complex_coordinate}), we have
\be
\Omega_0 
= \frac{2+r^2}{2}{\bf 1}_2 + \frac{r^2}{2}
\left(\begin{array}{cc}
- \cos 2\xi & e^{i(\lambda-\eta)}  \sin2\xi\\
e^{-i(\lambda-\eta)} \sin2\xi & \cos2\xi
\end{array}
\right).
\ee
\begin{figure}[t]
\begin{center}
\begin{tabular}{cc}
\includegraphics[height=4.8cm]{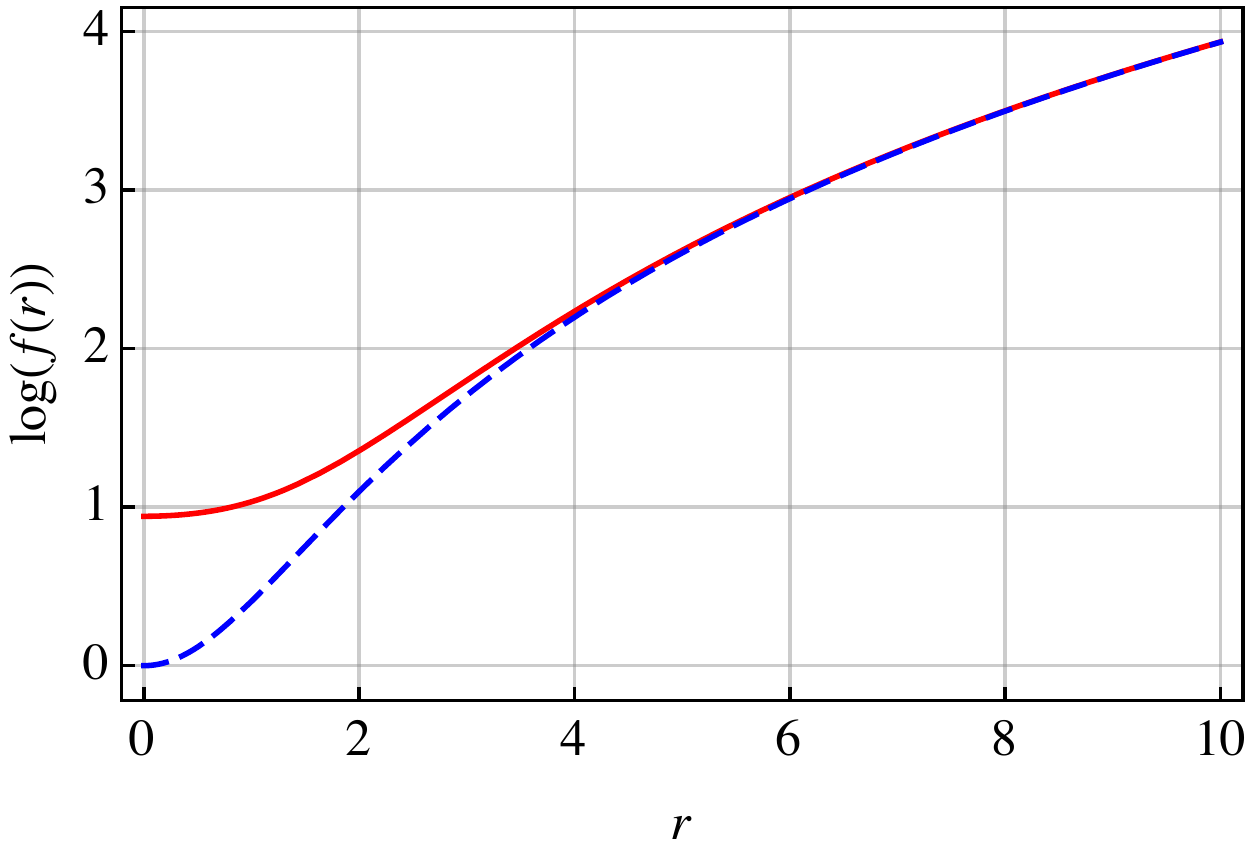} &
\includegraphics[height=4.8cm]{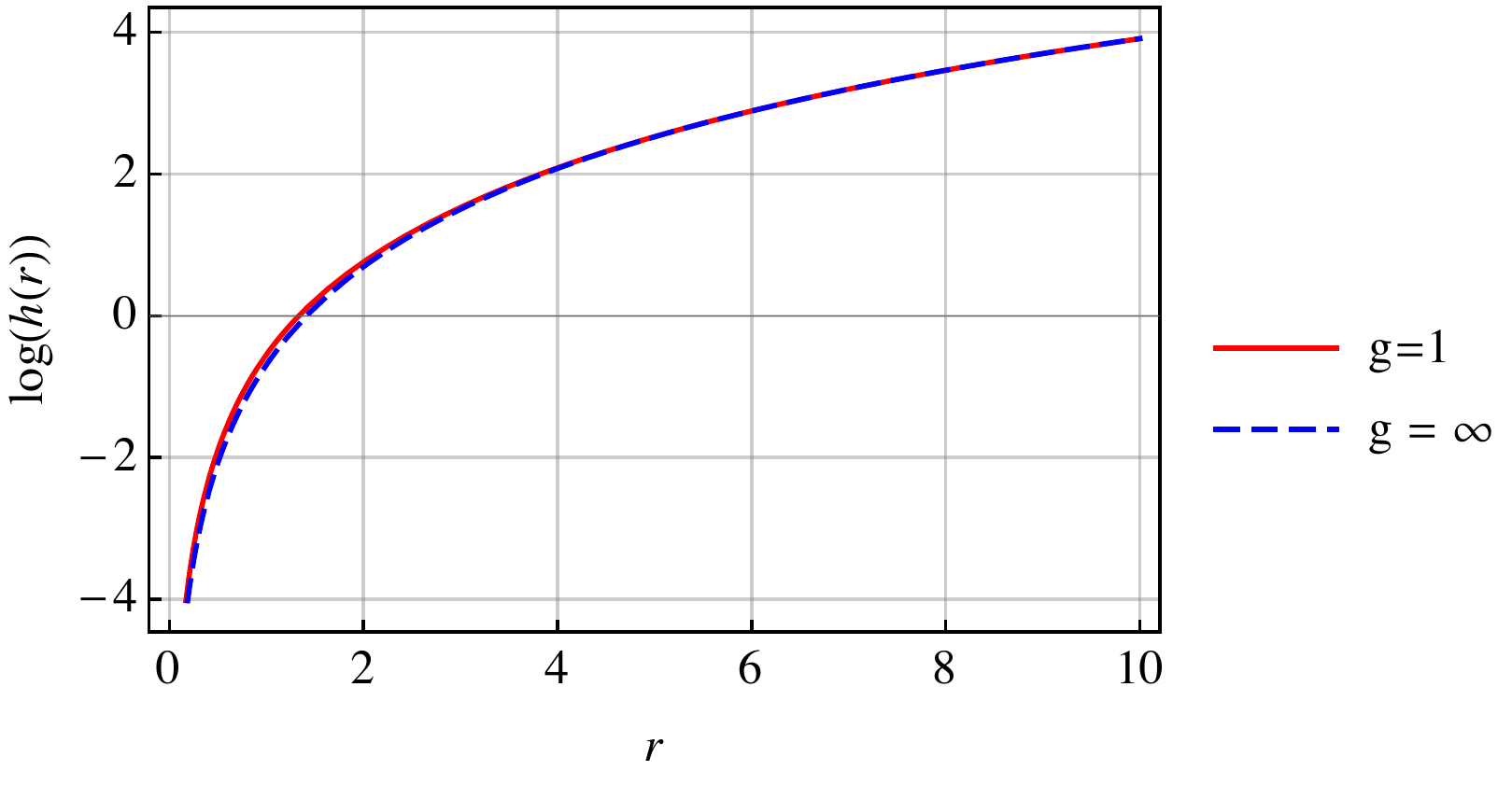}
\end{tabular}
\caption{Profile functions $\log f(r)$ and $\log h(r)$ are shown with
$cg^2=1,\infty$ for $a=1$.}
\label{fig:config_na_inst}
\ \\\ \\
\includegraphics[width=10cm]{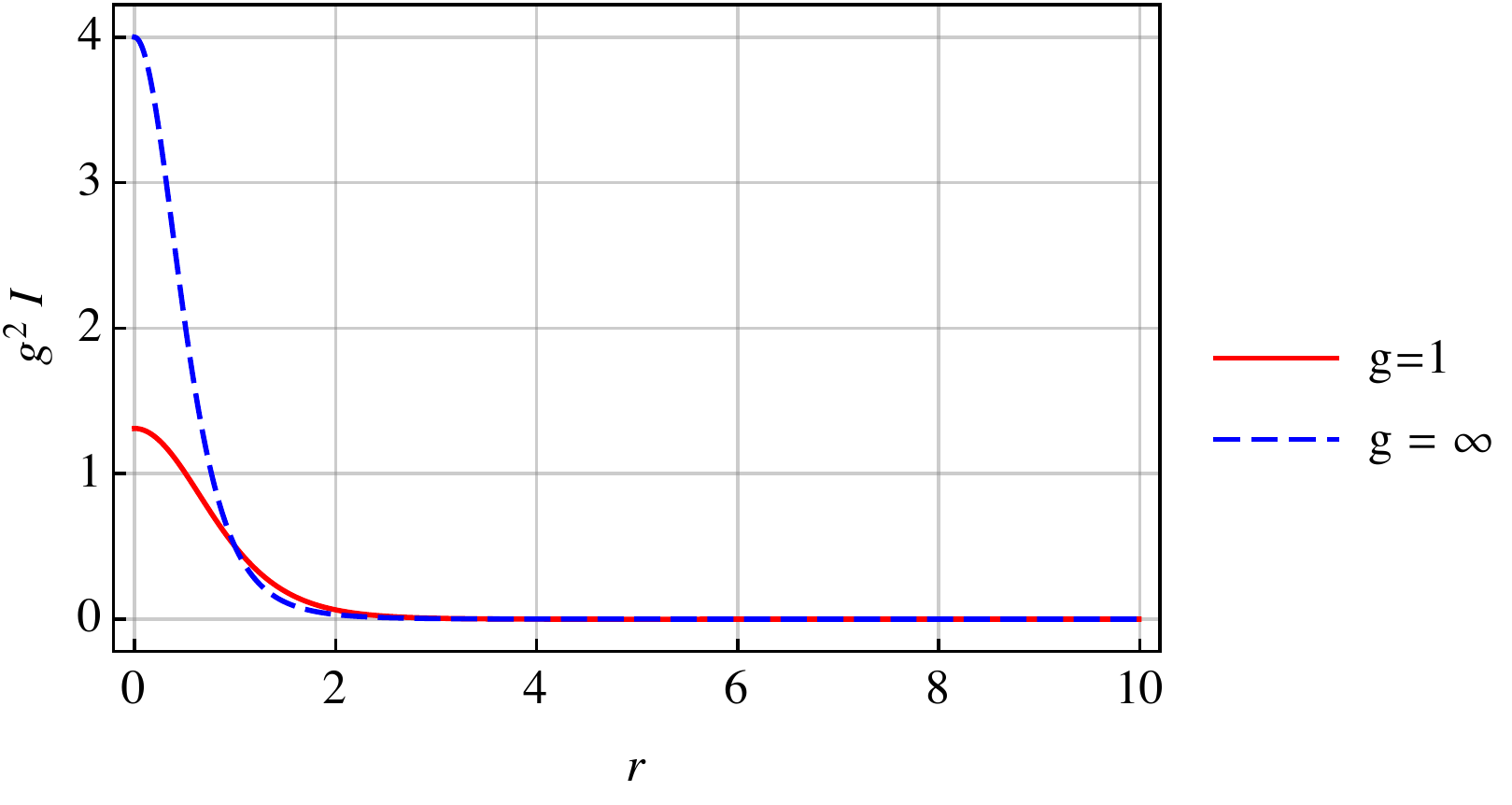} 
\caption{Instanton charge densities $g^2{\cal I}$ are shown with
$c g^2 =1,\infty$ for $a=1$.}
\label{fig:inst_density_na}
\end{center}
\end{figure}
Since $\Omega$ should asymptotically close to $\Omega_0$, we make an Ansatz for $\Omega$
\be
\Omega = 
f(r) {\bf 1}_2 + h(r)
\left(\begin{array}{cc}
- \cos 2\xi & e^{i(\lambda-\eta)}  \sin2\xi\\
e^{-i(\lambda-\eta)} \sin2\xi & \cos2\xi
\end{array}
\right),
\ee
where we impose
\be
f(r) \to \frac{2+r^2}{2},\quad 
h(r) \to \frac{r^2}{2},\qquad (r \to \infty).
\ee
Plugging the above Ansatz into the master equation (\ref{master_vvi}), we get the
following two ordinary differential equations
\be
f'' + \frac{3}{r}f' - \frac{1}{r^2}\frac{8 h^2}{f+h} - \frac{f f'{}^2 - 2 hf'h'+fh'{}^2}{f^2-h^2} 
+\frac{cg^2}{2}\left(2+r^2-2f\right) = 0,\\
h'' + \frac{3}{r}h' - \frac{1}{r^2}\frac{8 fh}{f+h} + \frac{h f'{}^2 - 2 ff'h'+hh'{}^2}{f^2-h^2} 
+\frac{cg^2}{2}\left(r^2-2h\right) = 0.
\ee
A numerical solution is shown in Fig.~\ref{fig:config_na_inst}.
The profile functions at the finite gauge coupling are slightly different only near the origin
from those at the infinite gauge coupling limit.
We also plot the instant charge density $g^2{\cal I}$ in Fig.~\ref{fig:inst_density_na}.
The density $g^2{\cal I}$ with $g=1$ becomes slightly smaller than that with $g =\infty$.
The peak of the density with $g=1$ is just 1/3 of 
that with $g=\infty$, which is quite different from the Abelian
case seen in the previous subsection.

\subsection{Seiberg-like duality}
\begin{figure}[t]
\begin{center}
\includegraphics[width=10cm]{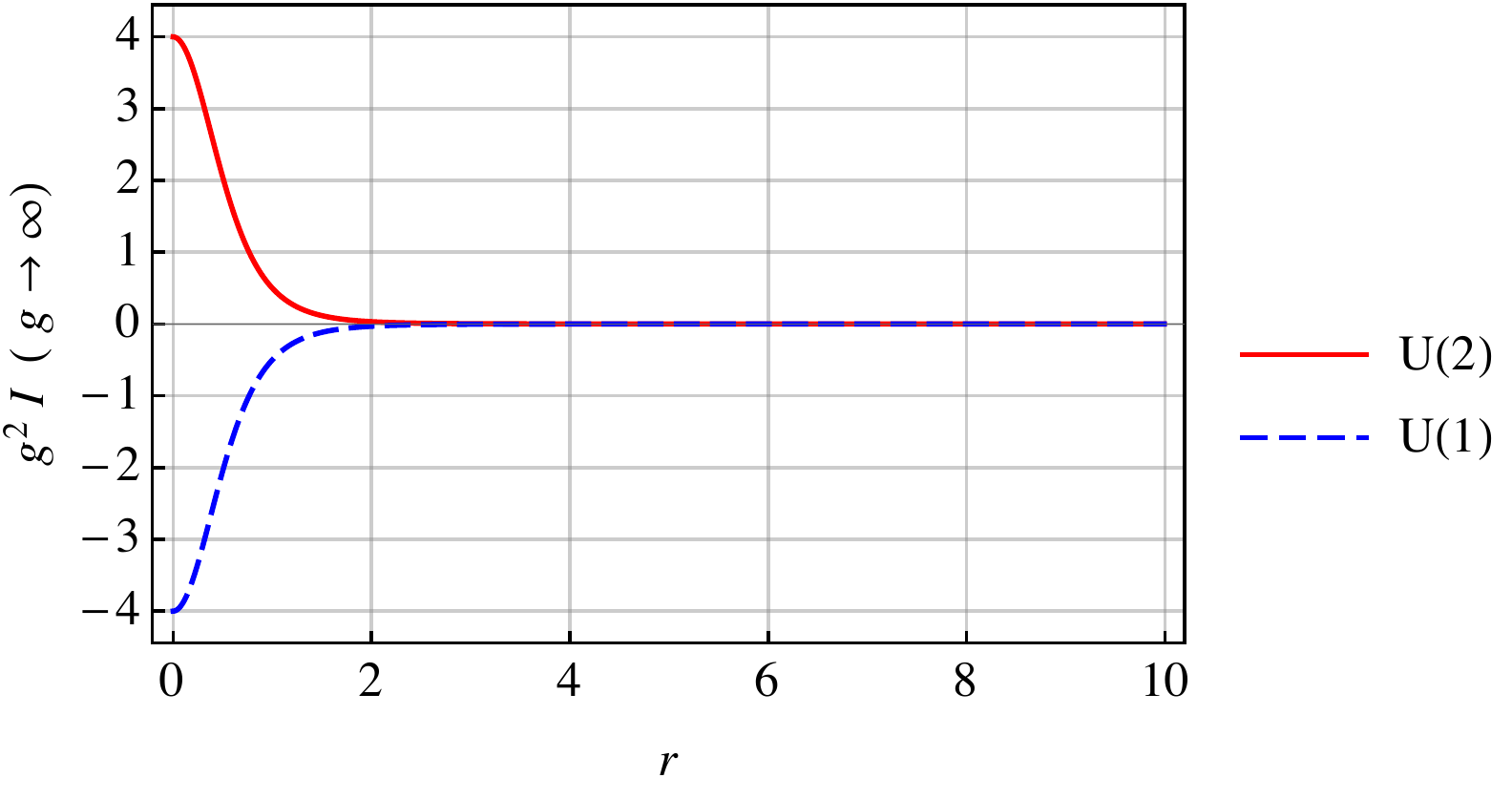} 
\caption{Instanton charge densities $g^2{\cal I}$ with
$c g^2 =\infty$ for $a=1$.}
\label{fig:instanton_charge_ab_nab_ginfty}
\ \\\ \\
\includegraphics[width=10cm]{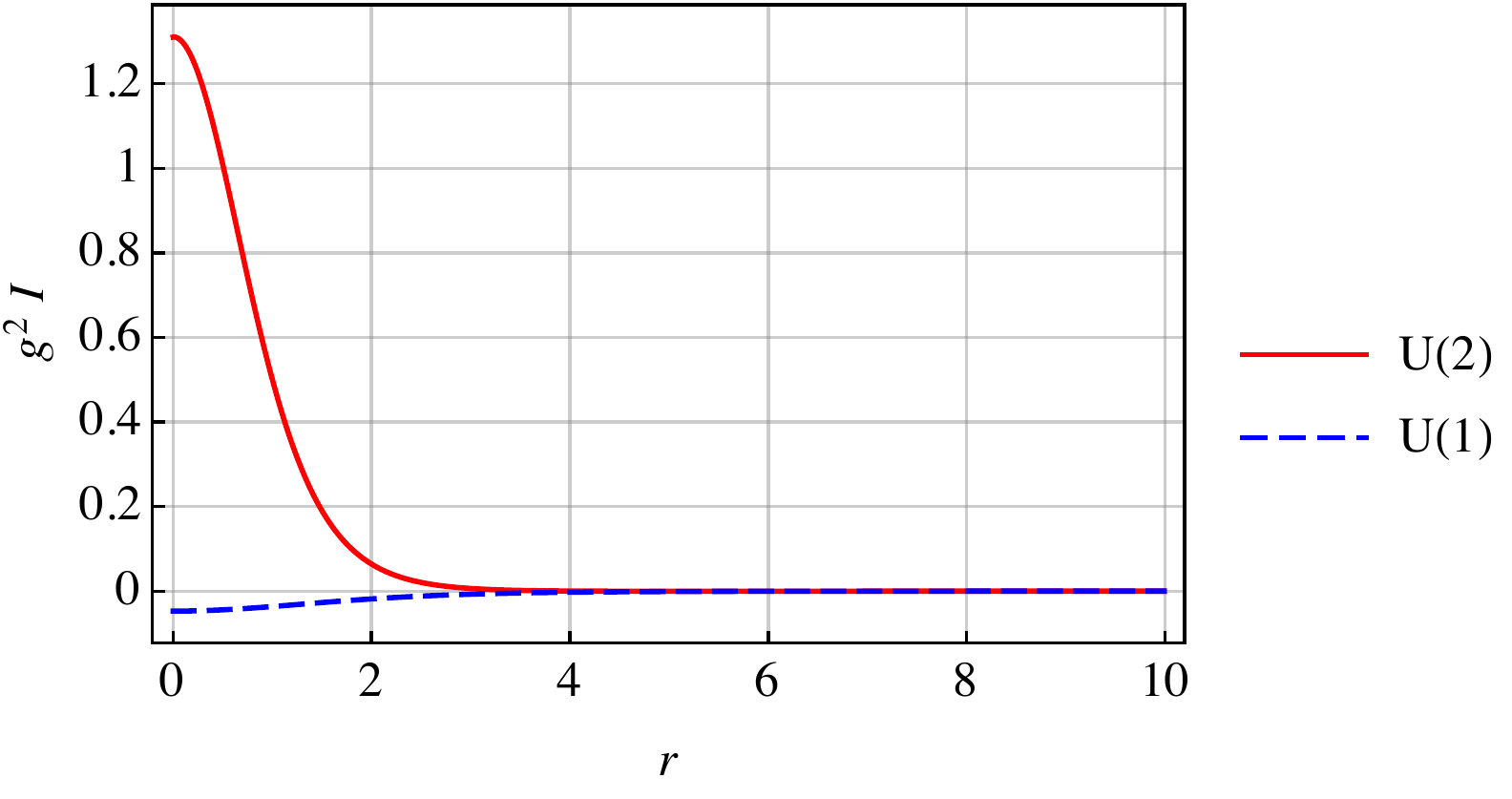} 
\caption{Instanton charge densities $g^2{\cal I}$ with
$c g^2 =1$ for $a=1$.}
\label{fig:instanton_charge_ab_nab_g1}
\end{center}
\end{figure}

In Figs.~\ref{fig:instanton_charge_ab_nab_ginfty} and \ref{fig:instanton_charge_ab_nab_g1}, 
we compare the instanton densities  $g^2{\cal I}$ of
the Abelian  and non-Abelian models for the finite gauge coupling $g=1$ and $g=\infty$.
Due to the duality $\NC \leftrightarrow \NF-\NC$ 
shown in Appendix \ref{sec:duality}, 
Fig.~\ref{fig:instanton_charge_ab_nab_ginfty} 
clearly shows that 
the instanton charge densities at the
infinite gauge coupling limit obey the exact relation 
$g^2{\cal I}_{\NC=2,\NF=3} = -g^2{\cal I}_{\NC=2,\NF=3}$. 

\section{Summary and Discussion}\label{sec:summary}
We have studied 1/4 BPS equations and 
have constructed semi-local 
fractional instantons of codimension four 
in $U(\NC)$ gauge theories with $\NF$ scalar fields 
in Euclidean four dimensions 
or the corresponding 
${\mathbb C}P^{\NF-1}$ and Grassmann sigma models 
in strong gauge coupling limit. 
In the sigma model limit, we have presented exact solutions,
and for finite gauge coupling we have given 
numerical solutions.
They have divergent energy in systems with infinite volume $R\to \infty$ as global solitons.
We find that they carry fractional  instanton charge 
$(-)1/2$ in non-Abelian (Abelian) gauge theories,
and that the Seiberg-like duality changes the sign of 
the instanton charge.

The topological origin of the fractionality 
of the topological charge is yet to be clarified.
Only when the spatial boundary is mapped to 
one point on the target space, 
the topological charge is integer. 
One known origin of the fractional topological charges 
is due to the presence of suitable potential term. 
For instance, 
a lump (sigma model instanton) in 
the ${\mathbb C}P^{N-1}$ model 
characterized by $\pi_2$ 
is split into $N$  fractional ($1/N$) lumps \cite{Schroers:1995he}, 
and  
a Skyrmion in the Skyrme model 
characterized by $\pi_3$ is split into 
two half Skyrmions 
\cite{Gudnason:2015nxa} 
in the presence of potential terms.
The other known example is given by
twisted boundary conditions 
along a compactified direction. 
There appear 
$1/N$ fractional lumps (sigma model instantons) 
of  $\pi_2$ in 
the ${\mathbb C}P^{N-1}$ model on ${\mathbb R}^1 \times S^1$ 
\cite{Eto:2004rz,Dunne:2012ae},   
fractional Grassmann lumps \cite{Eto:2006mz} 
on ${\mathbb R}^1 \times S^1$, 
 fractional instantons of $\pi_{N-1}$ 
in the $O(N)$ model 
on ${\mathbb R}^{N-2} \times S^1$
\cite{Nitta:2014vpa},  
and $1/N$ fractional instantons 
 of $\pi_3$ in the $SU(N)$
principal chiral model on ${\mathbb R}^2 \times S^1$
\cite{Nitta:2015tua}.\footnote{ 
The latter case is currently 
extensively studied in application to resurgence 
of field theory \cite{Dunne:2012ae}.
}
These two cases stabilize fractional charges 
by either the potential term or the boundary conditions.
Our new solitons are stabilized by different mechanism,
namely by the topological charges of vortices.
A common point for these three cases is that 
fractional solitons have additional 
topological charges of one less dimensions.
In our case it is the Hopf charge in Eq.~(\ref{eq:Hopf}).

A 1/2 BPS non-Abelian vortex in 
the $U(N)$ gauge theory has 
the ${\mathbb C}P^{N-1}$ moduli space 
\cite{Hanany:2003hp,Auzzi:2003fs}. 
In 5+1 dimensions a vortex has 
3+1 dimensional world-volume.  
If we consider our instanton solution in the ${\mathbb C}P^{N-1}$ model on the Euclid world-volume of the vortex, we have an instanton of codimension four inside 
a vortex of codimension two. This object of codimension six may be 1/8 BPS state in Euclidean six dimensional theory 
with eight supercharges \cite{Eto:2005sw}.

As the other coset spaces that admits the same semilocal 
instantons, we may consider the followings:
\beq
&& \pi_4 \left( \frac{SO(2N)}{SU(N) \times U(1)} \right)
= \pi_3 \left( SU(N) \times U(1) \right) = {\mathbb Z},\\
&& \pi_4 \left( \frac{USp(2N)}{SU(2) \times U(1)} \right)
= \pi_3 \left( SU(N) \times U(1) \right) = {\mathbb Z}.
\eeq
The nonlinear sigma models with these target spaces
 can be constructed 
in supersymmetric gauge theories with 
suitable superpotentials
in the case of four supercharges \cite{Higashijima:1999ki}. 
These sigma model also appear as the effective theory on a 
non-Abelian vortex 
in $G=SO, USp$ gauge theories \cite{Eto:2008yi}.

It is interesting to take quantum effects into account, although,  in this paper, we have focused our attention
to the classical aspects of the fractional semilocal instantons.
We have taken the common gauge coupling constant $g$ for the $U(1)$ and $SU(\NC)$ parts for usefulness.
The overall $U(1)$ would be free in the IR for the five dimensions, so 
we may need to solve the equations of motion with different gauge couplings for the $U(1)$ and $SU(\NC)$ parts. 
It would modify the solutions obtained in this paper
but the qualitative features like topological charges and masses may not be affected.

\section*{Acknowledgments}
 We thank Keisuke Ohashi and Toshiaki Fujimori for collaborations in related works. 
This work is supported by the MEXT-Supported Program for the Strategic
Research Foundation at Private Universities ``Topological Science''
(Grant No.~S1511006).
The work of M. E. is supported in part by
JSPS Grant-in-Aid for Scientic Research (KAKENHI Grant No.~26800119).
The work of M.~N.~is supported in part by a Grant-in-Aid for
Scientific Research on Innovative Areas ``Topological Materials
Science'' (KAKENHI Grant No.~15H05855) and ``Nuclear Matter in Neutron
Stars Investigated by Experiments and Astronomical Observations''
(KAKENHI Grant No.~15H00841) from the the Ministry of Education,
Culture, Sports, Science (MEXT) of Japan. The work of M.~N.~is also
supported in part by the Japan Society for the Promotion of Science
(JSPS) Grant-in-Aid for Scientific Research (KAKENHI Grant
No.~25400268).

\appendix
\section{Topological Charges in Seiberg-like Duality}\label{sec:duality}
In this section, we discuss the transformation of the 
topological charge under the Seiberg-like duality.
We show that the sign of the topological is flipped.

Let us consider
the strong gauge coupling limit 
$g^2 \rightarrow \infty$.
In this case, there is a duality between the theory 
with $(N_{\rm F},N_{\rm C})$ and with 
$(N_{\rm F},\tilde{N}_{\rm C} \equiv N_{\rm F}-N_{\rm C})$.
We introduce the dual scalar fields $\tilde H$ 
in the form of an $\tilde{N}_{\rm C} \times \NF$ matrix.
The scalar fields satisfy the constraints
\be
H H^\dagger = c{\bf 1}_{N_{\rm C}},\quad
\tilde H \tilde H^\dagger = c{\bf 1}_{\tilde N_{\rm C}}.
\ee
Then, there is the following relation between 
the original fields $H$ and the dual field $\tilde H$:
\be
H \tilde H^\dagger = 0\quad{\rm or}\quad
H^\dagger H + \tilde H^\dagger \tilde H = c{\bf 1}_{N_{\rm F}}.
\ee
The tension of the instanton is given by
\be
g^2 T_{\rm Instanton} = \int d^4x\ \frac{1}{2}{\rm Tr}
\left(\varepsilon^{MNKL}F_{MN} F_{KL}\right),
\label{eq:a3}
\ee
where the gauge field is expressed as
\be
W_M = \frac{i}{c}\partial_MH H^\dagger,\quad
\tilde W_M = \frac{i}{c}\partial_M\tilde H \tilde H^\dagger.
\ee
The field strength is then expressed as
\be
F_{MN} &=&
\partial_MW_N - \partial_NW_M
+ i\left[W_M,W_N\right]\nonumber\\
&=& \frac{i}{c}\left[
\partial_M\left\{\partial_NHH^\dagger\right\}
- \partial_N\left\{\partial_MHH^\dagger\right\}
\right]
- \frac{i}{c^2}\left[
\partial_MHH^\dagger,
\partial_NHH^\dagger
\right]
\nonumber\\
&=& -\frac{2i}{c}\left[
\partial_{[M}H\partial_{N]}H^\dagger
+ \frac{1}{c}\partial_{[M}HH^\dagger \partial_{N]}HH^\dagger
\right]\nonumber\\
&=& -\frac{2i}{c}\left[
\partial_{[M}H\partial_{N]}H^\dagger
- \frac{1}{c}\partial_{[M}HH^\dagger H\partial_{N]}H^\dagger
\right]\nonumber\\
&=& -\frac{2i}{c}\left[
\partial_{[M}H\partial_{N]}H^\dagger
- \frac{1}{c}\partial_{[M}H
\left( c{\bf 1}_{N_{\rm F}} - \tilde H^\dagger \tilde H \right)
\partial_{N]}H^\dagger
\right]\nonumber\\
&=& -\frac{2i}{c^2}\partial_{[M}H\tilde H^\dagger
\tilde H\partial_{N]}H^\dagger\nonumber\\
&=& \frac{2i}{c^2}\partial_{[M}H\tilde H^\dagger
\partial_{N]}\tilde HH^\dagger.
\label{eq:dual}
\ee
where we have defined
\be
\partial_{[M}X\partial_{N]}Y \equiv \frac{1}{2}
\left[\partial_MX\partial_NY - \partial_NX\partial_MY\right].
\ee
Plugging equation (\ref{eq:dual}) into the equation (\ref{eq:a3}), we get
\be
{\rm Tr}\left(\varepsilon^{MNKL}F_{MN} F_{KL}\right)
&=& - \frac{4}{c^4}{\rm Tr}\left[\varepsilon^{MNKL}
\partial_{[M}H\tilde H^\dagger
\partial_{N]}\tilde HH^\dagger
\partial_{[K}H\tilde H^\dagger
\partial_{L]}\tilde HH^\dagger
\right]\nonumber\\
&=& - \frac{4}{c^4}{\rm Tr}\left[\varepsilon^{MNKL}
\partial_{M}H\tilde H^\dagger
\partial_{N}\tilde HH^\dagger
\partial_{K}H\tilde H^\dagger
\partial_{L}\tilde HH^\dagger
\right]\nonumber\\
&=& - \frac{4}{c^4}{\rm Tr}\left[\varepsilon^{MNKL}
\partial_{N}\tilde HH^\dagger
\partial_{K}H\tilde H^\dagger
\partial_{L}\tilde HH^\dagger
\partial_{M}H\tilde H^\dagger
\right]\nonumber\\
&=& \frac{4}{c^4}{\rm Tr}\left[\varepsilon^{MNKL}
\partial_{M}\tilde HH^\dagger
\partial_{N}H\tilde H^\dagger
\partial_{K}\tilde HH^\dagger
\partial_{L}H\tilde H^\dagger
\right]\nonumber\\
&=& - {\rm Tr}\left(\varepsilon^{MNKL}\tilde F_{MN} \tilde F_{KL}\right)
\ee
Note that $\tilde F_{MN}$ is not the electric-magnetic 
dual of $F_{MN}$
but is the field strength of $\tilde W_M$.
We thus have found the relation
\be
T_{\rm Instanton} + \tilde T_{\rm Instanton} = 0
\ee
implying that the instanton charge is flipped 
under the duality.

The same relation between the monopole charges in 
the original and dual theories.
In the dimensional reduction, the instanton charge 
can be written as
\be
{\rm Tr}\left(\varepsilon^{MNKL}F_{MN}F_{KL}\right)
&=& 4{\rm Tr}\left(\varepsilon^{mnk4}F_{mn}F_{k4}\right)\nonumber\\
&\rightarrow& 4{\rm Tr}\left(\varepsilon^{mnk}F_{mn}{\cal D}_k\Xi\right)\nonumber\\
&=& 8{\rm Tr}\partial_k\left(^*\!F^k\Xi\right),
\ee
where we have defined the adjoint scalar field 
$\Xi(x^m) = W_4 $ and have used the Bianchi identity.
Therefore, we have 
\be
T_{\rm Instanton} &\rightarrow& T_{\rm Monopole},
\ee
and the relation 
\be
T_{\rm Monopole} + \tilde T_{\rm Monopole} = 0.
\ee

\newcommand{\J}[4]{{\sl #1} {\bf #2} (#3) #4}
\newcommand{\andJ}[3]{{\bf #1} (#2) #3}
\newcommand{\AP}{Ann.\ Phys.\ (N.Y.)}
\newcommand{\MPL}{Mod.\ Phys.\ Lett.}
\newcommand{\NP}{Nucl.\ Phys.}
\newcommand{\PL}{Phys.\ Lett.}
\newcommand{\PR}{ Phys.\ Rev.}
\newcommand{\PRL}{Phys.\ Rev.\ Lett.}
\newcommand{\PTP}{Prog.\ Theor.\ Phys.}
\newcommand{\hep}[1]{{\tt hep-th/{#1}}}

\end{document}